\documentstyle[12pt]{article}
\begin{document}
\def\be{\begin{eqnarray}}
\def\en{\end{eqnarray}}
\def\non{\nonumber}
\def\la{\langle}
\def\ra{\rangle}
\def\ep{\varepsilon}
\def\u{{\mu}_{\rm fact}}
\def\ums{{\mu}_{\rm MS}}
\def\lsim{ {\ \lower-1.2pt\vbox{\hbox{\rlap{$<$}\lower5pt\vbox{\hbox{$\sim$}
}}}\ } }
\def\gsim{ {\ \lower-1.2pt\vbox{\hbox{\rlap{$>$}\lower5pt\vbox{\hbox{$\sim$}
}}}\ } }
\def\pr{{\sl Phys. Rev.}~}
\def\prl{{\sl Phys. Rev. Lett.}~}
\def\pl{{\sl Phys. Lett.}~}
\def\np{{\sl Nucl. Phys.}~}
\def\zp{{\sl Z. Phys.}~}

\font\el=cmbx10 scaled \magstep2
{\obeylines
\hfill IP-ASTP-17-95
\hfill November, 1995
\hfill (Revised)}

\vskip 1.5 cm

\centerline{\large\bf Polarized Parton Distribution Functions Reexamined}
\medskip
\bigskip
\medskip
\centerline{\bf Hai-Yang Cheng and Hung Hsiang Liu}
\medskip
\centerline{ Institute of Physics, Academia Sinica}
\centerline{Taipei, Taiwan 115, Republic of China}
\bigskip
\centerline{\bf Chung-Yi Wu}
\medskip
\centerline{Department of Physics, National Cheng Kung University}
\centerline{Tainan, Taiwan 701, Republic of China}
\bigskip
\bigskip
\bigskip
\centerline{\bf Abstract}
\bigskip
{\small
   The hard-gluonic contribution to the first moment
of the polarized proton structure function $g_1^p(x)$ is dependent of
the factorization convention chosen in defining the quark spin density
and the hard cross section for photon-gluon scattering.
Two extremes of interest, namely
gauge-invariant and chiral-invariant factorization schemes,
are considered. We show that in order to satisfy the positivity
constraint for sea and gluon polarizations, the polarized valence quark
distributions should fully account for the observed $g_1^p(x)$ at $x\gsim
0.2\,$. This together with the first-moment and perturbative QCD constraints
puts a pertinent restriction on the shape of $\Delta u_v(x)$ and $\Delta
d_v(x)$. The spin-dependent sea distribution in the gauge-invariant
factorization scheme
is extracted from the data of $g_1^p(x)$. It is shown in the chiral invariant
scheme that it is possible to interpret the $g_1^p(x)$ data with anomalous
gluonic contributions, yet a best least $\chi^2$
fit to the data implies a gluon spin distribution which violates the
positivity condition $|\Delta G(x)|\leq G(x)$.
We then propose a more realistic set of parton spin distributions with sea
polarization and with a moderate value of $\Delta G$. The polarized parton
distributions in this work are presented in the next-to-leading order of QCD
at the scale $Q^2=10\,{\rm GeV}^2\,$.
Predictions for the polarized structure functions $g_1^n(x)$ of the neutron
and $g_1^d(x)$ of the deuteron are given.

}

\pagebreak

\noindent{\it\bf I. Introduction}

  In the last few years we have witnessed a remarkable progress in the study
of polarized hadron structure functions and the related proton spin issue.
Experimentally, new measurements of the longitudinal spin-dependent
structure functions on various targets in polarized deep inelastic
lepton-hadron scattering became available. The polarized structure functions
$g_1^p(x)$ of the proton [1,2], $g_1^n(x)$ of the neutron [3], and $g_1^d(x)$
of the deuteron [4,5] have been measured recently. The
original EMC experiment on $g_1^p(x)$ [6],
which has triggered a great deal of interest in the proton spin structure, is
confirmed by the new high-statistics experimental data. Theoretically, a
direct first-principles lattice QCD calculation of the proton matrix elements
of the axial vector current, which is free of the $\eta'$ and related problems
encountered before [7], is also available very recently [8,9]. The calculated
quark spin is consistent with experiment. It is also evident from the lattice
calculation that it is the disconnected diagram, which is presumably
dominated by the axial anomaly, that explains why the total spin carried by
the quarks in a polarized proton is smaller than naively expected.

     In spite of the aforementioned progress, the extraction of
spin-dependent parton distribution functions, especially for sea quarks and
gluons, from the measured polarized hadron structure functions remains largely
ambiguous and controversial. One main issue has to do with
the debate of whether or not gluons contribute to $\Gamma_1^p$, the first
moment of $g_1^p(x)$.
Depending on the interpretation on the discrepancy between
experiment and the naive expectation for $\Gamma_1^p$ (i.e.,
the Ellis-Jaffe sum rule [10]), two different sets of polarized parton
distributions are often presented in the literature in the following way.
First, one makes some parametrizations for spin-dependent parton densities
based on some plausible (model) constraints. Then fitting these
parametrizations to the data of $g_1^p(x)$ etc., one obtains (i) a best fit of
$\Delta u(x),~\Delta d(x)$ and $\Delta s(x)$ at fixed $Q_0^2$ by assuming
$\Delta G(x,Q^2_0)=0$, or (ii) a best fit of $\Delta G(x)$ and the polarized
valence distributions $\Delta u_v(x),~\Delta d_v(x)$ with no sea polarization.

    However, most of the parton spin densities presented in the literature
are problematic. First, model-independent QCD constraints on the valence
spin densities $\Delta u_v(x)$ and $\Delta d_v(x)$ at $x\to 1$ are not
respected
in many existing parametrizations. Second, most authors fail to employ a
correct kernel $\Delta\sigma^{\gamma G}(x)$, the hard cross section for
photon-gluon scattering, to evaluate the gluonic contribution to the
proton structure function $g_1^p(x)$. As we are going to stress in Sec. 2,
whether or not gluons contribute to $\Gamma_1^p$ is purely factorization
dependent [11]. Once a factorization scheme is chosen, the ``hard" kernel
is completely fixed up to the factorization scale $\u$. A determination of
parton spin distributions using any other kernels, for instance the delta
kernel, is certainly not trustworthy.

     In the present paper we shall give a critical analysis of the
polarized parton distributions. We first give a brief overview in Section II
on the role of the hard-gluonic contribution to the first moment of the
polarized proton structure function. Based on the gauge-invariant and
chiral-invariant factorization schemes, we then proceed to extract parton spin
distributions from the $g_1^p(x)$ data in Sections III and IV respectively.
Sections V and VI contain discussions and conclusions.

\vskip 0.3 cm
\noindent {\bf II.~~Framework}
\vskip 0.3 cm
   The sea-quark or anomalous gluonic interpretation for the violation of
the Ellis-Jaffe sum rule depends on the factorization scheme defined
for the quark spin density and
the cross section for photon-gluon scattering. Much of the factorization
scheme dependence and the related issues are already addressed by Bodwin and
Qiu [11]. To set up the notation and the results necessary for our
purposes, we will recapitulate the main points in Ref.[11].

   The general expression of the proton structure function in the presence
of QCD corrections to order $\alpha_s$ is
\be
g_1^p(x,Q^2) &=& {1\over 2}\sum_{i=1}^{n_f}e^2_i\int^1_x{dy\over y}\Bigg\{
\Delta q_i(y,Q^2)\left[\delta\left(1-{x\over y}\right)+{\alpha_s(Q^2)\over
2\pi}\Delta f_q\left({x\over y}\right)\right]   \non \\
&-& {\alpha_s(Q^2)\over 2\pi}\Delta\sigma^{\rm hard}
\left({x\over y}\right)\Delta G(y,Q^2)\Bigg\},
\en
where $\Delta f_q$ depends on the regularization scheme chosen. Since the
unpolarized parton distributions are usually parametrized and fitted to data
in the $\overline{\rm MS}$
scheme, it is natural to adopt the same regularization scheme for polarized
parton distributions in which $\Delta f_q(x)=f_q(x)-{4\over 3}(1+x)$ and
(see e.g., [12])
\be
f_q(x) &=& {4\over 3}\Bigg[ (1+x^2)\left({\ln(1-x)\over 1-x}\right)_+-{3\over
2}{1\over (1-x)_+}-\left({1+x^2\over 1-x}\right)\ln x  \non \\
&& +3+2x-\left({9\over 2}+{\pi^2\over 3}\right)\delta(1-x)\Bigg],
\en
where the ``$+$" distribution is given by
\be
\int^1_0 g(x)\left( {f(x)\over 1-x}\right)_+dx=\int^1_0 f(x)\,{g(x)-g(1)\over
1-x}dx.
\en
The first moment of $f_q(x)$ and $\Delta f_q(x)$ is 0 and $-2$ respectively.
The parton spin densities in Eq.(1) are defined by
$\Delta q(x)=q^\uparrow(x)+\bar{q}^\uparrow(x)-q^\downarrow(x)-\bar{q}^
\downarrow(x)$ and $\Delta G(x)=G^\uparrow(x)-G^\downarrow(x)$. It is known
that a direct calculation of the polarized photon-gluon scattering box
diagram indicates that $\Delta\sigma(x)$ has collinear and infrared
singularities when $m^2=p^2=0$, where $m$ is the quark mass and $p^2$ is the
four-momentum squared of the gluon. Depending on
the choice of the soft cutoff, one obtains
\footnote{It is known that $\int^1_0\Delta\sigma(x)dx=0$ in the $\overline{
\rm MS}$ scheme [11]. However, the ``hard" part of $\Delta\sigma(x)$ is
dependent of the factorization scheme chosen, as elucidated below. For this
reason, we shall discuss various soft cutoff schemes.}

   (i) $m^2=0$ and $p^2\neq 0$ [13]
\be
\Delta\sigma_{\rm CCM}(x)=\,(1-2x)\left(\ln{Q^2\over -p^2}+\ln{1\over x^2}-2
\right),
\en

   (ii) $m^2\neq 0$ and $p^2=0$ [14]
\be
\Delta\sigma_{\rm AR}(x)=\,(1-2x)\left(\ln{Q^2\over m^2}+\ln{1-x\over x}-1
\right)-2(1-x),
\en

   (iii) dimensional regularization [12]
\footnote{The last term $-2(1-x)$ in Eqs.(5) and (6) was neglected in the
original work of Altarelli and Ross [14] and of Ratcliffe [12] respectively.
It arises from chiral symmetry breaking due to the $m^2\neq 0$ cutoff in the
mass-regulator scheme and the violation of the identity $\{\gamma_\mu,
{}~\gamma_5\}=0$ in the dimensional regularization scheme when $\epsilon\neq
0$.
One may argue that this contribution is soft, for example, in the
mass-regulator scheme if $m^2<<\u^2$ and hence it does not contribute to
``hard"
$\Delta\sigma$.  However, the cancellation of the $\ln(Q^2/m^2)$ term, which
depends logarithmically on the soft cutoff, from different $x$ regions
is not reliable because chiral symmetry may be broken at some hadronic scale.}
\be
\Delta\sigma_{\rm R}(x)=\,(1-2x)\left({1\over\epsilon}+\gamma_{\rm E}+\ln
{Q^2\over 4\pi\ums^2}+\ln{1-x\over x}-1\right)-2(1-x),
\en
where $\ums$ is a regulated scale in the minimal-subtraction scheme.
For the first moment of $\Delta \sigma(x)$, it is easily seen that
\be
\int^1_0\Delta\sigma_{\rm CCM}(x)dx=1,~~~\int^1_0\Delta\sigma_{\rm AR}(x)dx=
\int^1_0\Delta\sigma_{\rm R}(x)dx=0.
\en
The result (7) can be understood as follows. Any term which is antisymmetric
under $x\to 1-x$, for instance terms proportional to const.$\times(1-2x)$,
makes no contribution to $\int^1_0\Delta\sigma(x)dx$, a consequence
of chiral symmetry or helicity conservation, recalling that the gluon splitting
function is of the form $\Delta P_{qG}(x)={1\over 2}(2x-1)$. However,
there is a chiral-symmetry-breaking term proportional to $(1-x)$ in the
mass-regulator and dimensional regularization schemes, which compensates
the hard contribution arising from the region $k^2_\perp\sim Q^2$, where
$k_\perp$ is the transverse momentum of the quark in the photon-gluon
box diagram.

   Now, in order to consider hard-gluonic contributions to $g_1^p(x)$
(by ``hard", we mean contributions with $k_\perp^2\gsim \u^2$), one
has to introduce a factorization scale $\u$ to subtract the unwanted soft
contribution, i.e., the contribution arising from the distribution of quarks
and antiquarks in a gluon:
\be
\Delta\sigma^{\rm hard}(x,Q^2/\u^2)=\,\Delta\sigma(x,Q^2)-\Delta\sigma^{\rm
soft}(x,\u^2).
\en
In practice, one makes an approximate expression for the box diagrams that is
valid for $k_\perp^2<<Q^2$ and then introduces an ultraviolet cutoff on the
integration variable $k_\perp$
to ensure that only the region $k^2_\perp\lsim\u^2$ contributes to the soft
part [11]. The choice of the regulator specifies the factorization convention.
When the ultraviolet cutoff is
gauge invariant, it breaks chiral symmetry due to the presence of the axial
anomaly and hence
makes a contribution to $\Delta\sigma^{\rm soft}$. Using the
dimensional regulator for the ultraviolet cutoff it follows that [11]
\be
\Delta\sigma_{\rm CCM}^{\rm soft}(x) &=& (1-2x)\left(-{1\over\epsilon}-\gamma_
{\rm E}+\ln{4\pi\ums^2\over -p^2}+\ln
{1\over x(1-x)}-1\right)+2(1-x),   \non \\
\Delta\sigma_{\rm AR}^{\rm soft}(x) &=& (1-2x)\left(-{1\over\epsilon}-\gamma_{
\rm E}+\ln{4\pi\ums^2\over m^2}\right),   \\
\Delta\sigma_{\rm R}^{\rm soft}(x) &=& 0,   \non
\en
for various soft cutoffs, and hence
\be
\int^1_0\Delta\sigma_{\rm CCM}^{\rm soft}(x)dx=1,~~~\int^1_0\Delta\sigma_{\rm
AR}^{\rm soft}(x)dx=\int^1_0\Delta\sigma_{\rm R}^{\rm soft}(x)dx=0.
\en
In the mass-regulator and dimensional-regulator schemes, the original
soft contributions in (5) and (6)
are canceled by the contribution from chiral symmetry breaking introduced by
the ultraviolet cutoff. Therefore, in the gauge-invariant factorization scheme
\be
\Delta\sigma^{\rm hard}(x)=\,(1-2x)\left(\ln{Q^2\over \u^2}+\ln{1-x\over x}-1
\right)-2(1-x),
\en
where $\u^2=4\pi\ums^2\exp(-\gamma_{\rm E}-1/\epsilon)$. It follows that
$\int^1_0\Delta\sigma^{\rm hard}(x)dx=0$. Note that $\Delta\sigma^{\rm
hard}(x)$ is independent of the choice of soft and ultraviolet regulators.
In this scheme, the quark spin has a gauge-invariant local operator
definition:
\be
s_\mu\Delta q=\la p|\bar{q}\gamma_\mu\gamma_5q|p\ra,
\en
where $s_\mu$ is the proton spin vector; it is $Q^2$ dependent because
of the nonvanishing two-loop anomalous dimension associated with the
flavor-singlet
quark operator. The fact that gluons do not contribute
to $\Gamma_1^p$ is in accordance with the OPE analysis in which only the
quark operator contributes to $\Gamma_1^p$ at the twist-2, spin-1 level [15]:
\be
\int^1_0 g_1^p(x)dx=\,{1\over 2}\left(1-{\alpha_s\over\pi}\right)\la
p^\uparrow|\sum e_q^2\bar{q}\gamma_\mu\gamma_5q|p^\uparrow\ra s^\mu=
{1\over 2}\left(1-{\alpha_s\over\pi}\right)\left({4\over
9}\Delta u+{1\over 9}\Delta d+{1\over 9}\Delta s\right),
\en
where $\Delta q\equiv\int^1_0\Delta q(x)dx$.

     By contrast, it is also possible to choose a chiral-invariant but
gauge-variant ultraviolet cutoff, so that [11]
\be
\Delta\tilde{\sigma}^{\rm hard}(x)=\,(1-2x)\left(\ln{Q^2\over \u^2+m^2-p^2
x(1-x)}+\ln{1-x\over x}-1\right)-(1-x)\,{2m^2-p^2x(1-2x)\over
\u^2+m^2-p^2x(1-x)}
\en
and $\int^1_0\Delta\tilde{\sigma}^{\rm hard}(x)dx=1$ for $\mu^2>>p^2,~
m^2$. In this chiral-invariant factorization scheme,
the quark spin distributions in a gluon are obtained by a direct cutoff
on the $k_\perp$ integration:
\be
\Delta q'^G(x,\u^2)=\int^{\u^2}_0d^2k_\perp\Delta q^G(x,k_\perp);
\en
that is, all the quarks with $k_\perp^2\lsim \u^2$ in the gluon distribution
are factored into the quark spin distribution. Contrary to the first scheme,
$\Delta q'\equiv\int^1_0\Delta q'(x)dx$ cannot be
written as a matrix element of a gauge-invariant local operator,
\footnote{Other main disparities between $\Delta q$ and $\Delta q'$
are as follows. (i) It is perhaps less known that [16] the spin-dependent
Altarelli-Parisi evolution equations apply directly only to the {\it
gauge-invariant} parton spin distributions. To evaluate the $Q^2$ evolution
of $\Delta q'(x)$ and $\Delta q'$, one has to first apply Eq.(36) for example.
(ii) In principle, $\Delta q'$ and $\Delta G$ have a simple partonic
definition: the former (latter) can be identified in one-jet (two-jet)
events in polarized deep inelastic scattering [13].}
but it is $Q^2$ independent as the gauge-variant ultraviolet cutoff in this
scheme does not flip helicity; it is thus close and parallel to the naive
intuition in the parton model that the
quark helicity is not affected by gluon emissions. Replacing $\Delta q(x)$
by $\Delta q'(x)$ and $\Delta\sigma^{\rm hard}(x)$ by $\Delta\tilde{\sigma}^{
\rm hard}(x)$ in Eq.(1), we find
\be
\int^1_0g_1^p(x)dx=\,{1\over 2}\left(1-{\alpha_s\over\pi}\right)\sum e_q^2(
\Delta q'-{\alpha_s\over 2\pi}\Delta G).
\en
Consequently, $\Delta q$ and $\Delta q'$ are related by
\be
\Delta q=\,\Delta q'-{\alpha_s\over 2\pi}\Delta G.
\en

    Finally, we notice that it is also possible to choose an intermediate
ultraviolet cutoff scheme which is neither gauge nor chiral invariant, so
in general $\Delta q=\,\Delta q'-\lambda{\alpha_s\over 2\pi}\Delta G$ for
arbitrary $\lambda$ ($\lambda=0$ and $\lambda=1$ corresponding to gauge- and
chiral-invariant factorization schemes, respectively) [17].
It is clear that the issue of whether or not
gluons contribute to $\Gamma_1^p$ is purely a matter of the factorization
scheme chosen in defining the quark spin density and the hard gluon-photon
scattering cross section; a change of the factorization convention merely
shifts the contribution of $\Delta q(x)$ and $\Delta\sigma^{\rm hard}(x)$
in such a way that the physical proton-photon cross section remains unchanged.
Though this controversy was resolved sometime ago by Bodwin and Qiu [11] (see
also Manohar [18], Bass and Thomas [16]), it is considerably
unfortunate that many of recent articles are still biased on the
anomalous-gluonic interpretation of the $g_1^p(x)$ data and that the work of
Bodwin and Qiu is either overlooked or not widely recognized and well
appreciated in the literature.

\vskip 0.3cm
\noindent {\bf III.~Polarized parton distributions in the gauge-invariant
factorization scheme}
\vskip 0.3cm
    This section is devoted to studying the spin-dependent valence and sea
distributions
based on the $g_1^p(x)$ data. We shall see that the positivity condition
$|\Delta s(x)|\leq s(x)$ due to the positivity of unpolarized parton
distribution puts a very useful constraint on the shape of the
polarized valence quark distributions. The presence of the gluon polarization
will affect the shape of $\Delta s(x)$, but not its first moment.

     To begin with, the combination of all EMC, SMC, E142 and E143 data for
$\Gamma_1^p$ together with the SU(3) parameters [19] $F+D=1.2573\pm 0.0028$
and $3F-D=0.579\pm 0.026$ yields [20]
\be
\Delta u=\,0.83\pm 0.03\,,~~~\Delta d=-0.43\pm 0.03\,,~~~\Delta s=-0.10\pm
0.03\,,
\en
and hence
\be
\Delta\Sigma\equiv \Delta u+\Delta d+\Delta s=\,0.31\pm 0.07\,.
\en
Decomposing $\Delta q$ into its valence and sea components $\Delta q=\Delta
q_v+\Delta q_s$, we shall follow Ref.[21] to assume that sea polarization
is SU(3) invariant,
i.e., $\Delta u_s=\Delta d_s=\Delta s$. This assumption is justified since it
leads to $\Delta u_v+\Delta d_v=0.60$ from Eq.(18), which is very close to
the naive expectation that $\Delta\Sigma=3F-D=0.579$ in the absence
of sea polarization. The sea polarization is also found to be SU(3)
symmetric within errors in the lattice calculation [8,9]. This is
understandable since the disconnected insertions (for a definition of
connected and disconnected insertions, see [8,9]), from which the sea-quark
polarization originates, are presumably dominated by the triangle diagram
and hence are independent of the light quark masses in the loop. (This effcet
is absent in unpolarized distributions). Therefore, for SU(3) symmetric sea
polarization, we obtain from (18) that
\be
\Delta u_v=\,0.93\,,~~~~\Delta d_v=-0.33\,.
\en
The Monte Carlo computation [8,9] shows that the magnitude of
valence quark polarizations arising from the connected diagram is close to
that given by (20).

   In terms of valence and sea spin distributions, Eq.(1) can be recast
to the form
\be
g_1^p(x,Q^2) &=& {1\over 2}\int^1_x{dy\over y}\Bigg\{
\left[{4\over 9}\Delta u_v(y,Q^2)+{1\over 9}\Delta d_v(y,Q^2)
+{2\over 3}\Delta s(y,Q^2)\right]   \\
&\times& \left[\delta\left(1-{x\over y}\right)+{\alpha_s(Q^2)\over 2\pi}
\Delta f_q\left({x\over y}\right)\right]
-{\alpha_s(Q^2)\over 6\pi}\,\Delta\sigma^{\rm hard}
\left({x\over y},{Q^2\over \u^2}\right)\Delta G(y,Q^2)\Bigg\}.   \non
\en
Recall that in the gauge-invariant factorization scheme, gluons
contribute to $g_1^p(x)$, but not to $\Gamma_1^p$. In general, both sea
quarks and gluons contribute to the polarized structure function, but we
will begin with the extreme case (i) $\Delta s(x)\neq 0$, $\Delta G(x)=0$.
Since the unpolarized sea distribution  is small at $x>0.2$, the positivity
constraint $|\Delta s(x)|\leq s(x)$ implies that the data of $g_1^p(x)$
at $x>0.2$ should be almost accounted for by $\Delta u_v(x)$ and
$\Delta d_v(x)$. Therefore, the shape of the spin-dependent valence quark
densities is nicely restricted by the measured $g_1^p(x)$ at $x>0.2\,$
together with the first-moment constraint (20) and the perturbative QCD
requirement [22] that valence quarks at $x=1$
remember the spin of the parent proton, i.e., $\Delta u_v(x)/u_v(x),~\Delta
d_v(x)/d_v(x)\to 1$ as $x\to 1$. In order to ensure the
validity of the positivity condition $|\Delta q_v(x)|\leq q_v(x)$, we choose
the MRS(A$'$) set [23] parametrized in the $\overline{\rm MS}$ scheme
at $Q^2=4\,{\rm GeV}^2$ as unpolarized valence parton distributions
\be
u_v(x,Q^2=4\,{\rm GeV}^2) &=& 2.26\,x^{-0.441}(1-x)^{3.96}(1-0.54\sqrt{x}
+4.65x),   \non \\
d_v(x,Q^2=4\,{\rm GeV}^2) &=& 0.279\,x^{-0.665}(1-x)^{4.46}(1+6.80\sqrt{x}
+1.93x).
\en
Accordingly, we must employ the same $\overline{\rm MS}$ scheme for polarized
parton distributions [see Eq.(1)] in order to apply the positivity constraint.
For the spin-dependent valence distributions we assume that they have the form
\be
\Delta q_v(x)=\,x^\alpha(1-x)^\beta(a+b\sqrt{x}+cx+dx^{1.5}),
\en
with $\alpha$ and $\beta$ given by Eq.(22). We find that an additional
term proportional to $x^{1.5}$ is needed in (23) in order to satisfy the
above three constraints.

   For the data of $g_1^p(x)$, we will use the SMC [1] and EMC [6] results,
both being measured at the mean value of $Q^2_0=10\,{\rm GeV}^2$. Following
the SMC analysis we have used the new $F_2(x)$ structure function measured by
NMC [24], which has a better
accuracy at low $x$, to update the EMC data (see Fig.~1). The best least
$\chi^2$ fit to $g_1^p(x)$ at $x\gsim 0.2$ by (23) is found to be
\footnote{It was assumed in Ref.[21] that $\Delta u_v(x)=\alpha(x) u_v(x)$,
$\Delta d_v(x)=\beta(x)d_v(x)$ with $\alpha(x),~\beta(x)\to 1$ as $x\to 1$ and
$\alpha(x),~\beta(x)\to 0$ as $x\to 0$. However, the constraint at $x=0$ is
not a consequence of QCD. In the present work we find that $\Delta u_v(x)/
u_v(x)=0.41$ and $\Delta d_v(x)/d_v(x)=-0.136$ at $x=0$. As a result, $|\Delta
q_v(x)|$ is usually larger than $|\Delta s(x)|$ even at very small $x$ (see
Fig.~5 below).}
\be
\Delta u_v(x,Q^2_0) &=& x^{-0.441}(1-x)^{3.96}(0.928+0.149\sqrt{x}-1.141x
+11.612x^{1.5}),   \non \\
\Delta d_v(x,Q^2_0) &=& x^{-0.665}(1-x)^{4.46}(-0.038-0.43\sqrt{x}-5.260x+8.443
x^{1.5}),
\en
at $Q^2_0=10\,{\rm GeV}^2$, which satisfies all aforementioned constraints.
\footnote{We have evoluted $q_v(x,Q^2)$ from $Q^2=4\,{\rm GeV}^2$ to $10\,{\rm
GeV}^2$ in order to compare with $\Delta q_v(x,Q^2_0)$.}
Since $\Delta d_v$ is negative while $\Delta d_v(x)$ is positive as $x\to 1$,
it means that the sign of $\Delta d_v(x)$ flips somewhere at $x=x_0$ [25].
We find that $x_0=0.496$ in our case.

   It is evident from Fig.~1 that a negative sea polarization is required to
explain the observed $g_1^p(x)$ at small $x$. Assuming $\Delta G(x,Q^2_0)=0$
at this moment, we find from (21), (24) and the data of
$g_1^p(x)$ that the polarized strange quark distribution is determined to be
\be
\Delta s(x,Q^2_0)=-x^{-1.17}(1-x)^{9.63}(0.013\sqrt{x}+0.862x-1.186x^{1.5}),
\en
with $\Delta s=-0.109$ and $\chi^2/{\rm d.o.f.}=12.24/22$,
where uses of $\u\sim 1$ GeV and $Q^2=Q^2_0=10\,{\rm GeV}^2$ have been made.
It is easily seen from Fig.~2 that the positivity
condition $|\Delta s(x)/s(x)|\leq 1$ is respected.

   To illustrate the importance of having a least $\chi^2$ fit of
$g_1^p(x)$ at $x\gsim 0.2$ by $\Delta u_v(x)$ and $\Delta d_v(x)$, let us
consider another parametrization as an example
\footnote{This parametrization is taken from Ref.[26] expect that we have made
a different normalization in order to satisfy the first-moment
constraint (20).}
\be
\Delta u_v(x) &=& 0.3588\,x^{-0.54}(1-x)^{3.64}(1+18.36x), \non \\
\Delta d_v(x) &=& -0.1559\,x^{-0.54}(1-x)^{4.64}(1+18.36x),
\en
with $\Delta u_v=0.93$ and $\Delta d_v=-0.33\,$. It is evident that, contrary
to Fig.~1, this parametrization gives a reasonable eye-fit to the data
(though $\chi^2/{\rm d.o.f.}=30/22$) even at small $x$, as
depicted in Fig.~3. One cannot tell if there is a truly
discrepancy between theory and experiment unless the first moment of
$g_1^p(x)$ is calculated and compared with data, i.e., $(\Gamma_1^p)_{\rm
theory}=0.176\pm 0.006$ versus $(\Gamma_1^p)_{\rm expt}=0.142\pm 0.008\pm
0.011$ [1]. Following the same procedure as before, we find that the
sea polarization necessary to fit the data violates the positivity
condition when $x>0.2\,$. This
example gives a nice demonstration that an eye-fit to the data can be quite
misleading. Therefore, we conclude that in order to satisfy the positivity
constraint due to sea polarization, valence quark spin densities should fully
accout for the observed $g_1^p(x)$ at $x\gsim 0.2$. As a consequence, a
deviation of theory from experiment for the polarized structure function
should manifest at small $x$.

  It has been argued that a bound on $\Delta s$, namely $|\Delta s|\leq
0.052^{+0.023}_{-0.052}$ [27], can be derived based on the information of
the behavior of $s(x)$ measured in deep inelastic neutrino
experiments and on the positivity constraint. However, this argument is quite
controversial [28]. We note that since the strange quark distribution
parametrized by  MRS(A$'$) [23] yields $\int^1_0 xs(x)dx=0.0182$ for the
strange sea momentum, it is consistent with the bound
$\int^1_0xs(x)dx\leq 0.048\pm 0.022$ extracted from the neutrino-nucleon
experiment [29]. Hence,
our $\Delta s(x)$ does satisfy all known constraints. More importantly, a sea
polarization of order $-0.11$ in the polarized proton is confirmed by lattice
calculations [8,9].

    In a realistic case, it is very unlikely that $\Delta G(x,Q^2)$ vanishes
at some scale $Q_0^2$ for all $x$. Even if $\Delta G(x,Q_0^2)=0$ at $Q^2=
Q_0^2$, it can be radiatively generated at $Q^2>Q_0^2$. In the absence of
any information on the shape and the magnitude of gluon polarization except
for the restriction $|\Delta G(x)/G(x)|\leq 1$, we first take
\be
\Delta G(x,Q^2_0)=\,2.5A_G(1-x)^{7.44},
\en
with $A_G=8.44$ and $\Delta G=2.5\,$, as an illustration. This parametrization
is taken from the set A of gluon distribution in
Ref.[25] but with a different normalization for our purpose.
We see from Fig.~4 that the effect of polarized gluons is to suppress
$g_1^p(x)$ at $x\lsim 0.01$ and enhance $g_1^p(x)$ at $0.01<x<0.15$ so that
the net contribution to $\Gamma_1^p$ vanishes; that is, hard gluons
contribute to $g_1^p(x)$ but not to $\Gamma_1^p$ in the gauge-invariant
factorization scheme. Since a realistic polarized gluon distribution ought
to have its first moment lie somewhere between 0 and 2.5\,, we take
\be
\Delta G(x,Q^2_0)=\,0.199x^{-1.17}(1-x)^{5.33}(0.03-1.71\sqrt{x}+3.01x+43.5
x^{1.5}),
\en
as determined below for case (iv). The first moment of this gluon spin density
is $\Delta G=0.5\,$. The presence of $\Delta G(x)$ will affect the shape of
$\Delta s(x)$ but not its first moment. Following the same extracting
procedure as before for $\Delta s(x)$, we find
\be
\Delta s(x,Q^2_0)=-x^{-1.17}(1-x)^{9.63}(0.014\sqrt{x}+0.865x-1.189x^{1.5}),
\en
with $\Delta s=-0.11$ and $\chi^2/{\rm d.o.f.}=11.75/22$.
The parametrizations (28) and (29) are regarded as the representative
spin-dependent parton distributions for case (ii), as exhibited in Fig.~5.

\vskip 0.4cm
\noindent{\bf IV.~Polarized parton distributions in the chiral-invariant
factorization scheme}
\vskip 0.4cm
   As elaborated on in Sec.~II, in the chiral-invariant factorization scheme
the quark spin $\Delta q'$ is $Q^2$ independent, and gluons
contribute to the first moment of the polarized proton structure function.
Since $\Delta q'=\Delta q+{\alpha_s\over 2\pi}\Delta G$ and $\Delta
s_v=0$, it is obvious that $\Delta q'_v=\Delta q_v$ for SU(3) symmetric
sea polarization. We shall follow Ref.[21] to assume that this is also true for
their $x$ dependence, i.e., $\Delta q'_v(x)=\Delta q_v(x)$. Since
$\Delta G(x)$ is also independent of the factorization chosen, we thus have
\be
g_1^p(x,Q^2) &=& {1\over 2}\int^1_x{dy\over y}\Bigg\{
\left[{4\over 9}\Delta u_v(y,Q^2)+{1\over 9}\Delta d_v(y,Q^2)
+{2\over 3}\Delta s'(y,Q^2)\right]   \\
&\times& \left[\delta\left(1-{x\over y}\right)+{\alpha_s(Q^2)\over 2\pi}
\Delta f_q\left({x\over y}\right)\right]
- {\alpha_s(Q^2)\over 6\pi}\,\Delta\tilde{\sigma}^{\rm
hard}\left({x\over y},{Q^2\over \u^2}\right)\Delta G(y,Q^2)\Bigg\},   \non
\en
with $\Delta\tilde{\sigma}^{\rm hard}$ being given by (14). We note that
several different expressions for the kernel have been employed in the
literature. For example, $\Delta\tilde{\sigma}(z)=\delta(1-z)$ was used by
Altarelli and Stirling [30], $\Delta\tilde{\sigma}(z)=(1-2z)\ln[(1-z)/z]$ by
Ellis et al. [31] and by Ross and Roberts [32]. However, as we have stressed
in Sec. II, a correct procedure of subtracting the soft contribution from
$\Delta\tilde{\sigma}(z)$ will yield a unique $\Delta\tilde{\sigma}^{\rm
hard}(z)$ up to the factorization scale $\u$, which is independent of the
choice of soft and ultraviolet regulators.

    We first discuss the extreme case, namely (iii) $\Delta s'(x)=0$
and $\Delta G(x)\neq 0$, which is just opposite to the other extreme case (i).
It has been advocated that [33,13] a total absence of sea polarization and an
anomalous gluonic contribution might offer an attractive and plausible
solution to the so-called ``proton spin crisis" by
accounting for the discrepancy between
the Ellis-Jaffe sum rule and experiment for $\Gamma_1^p$.
It follows from (18) that $\Delta\Sigma'=0.58$ with $\Delta s'=0$, consistent
with what expected from the relativistic quark model. To implement
a large $\Delta\Sigma'$ and a vanishing $\Delta s'$ demands a large gluon
polarization: $\Delta G=-(2\pi/\alpha_s)\Delta s=2.5$ at $Q^2=10\,{\rm
GeV}^2$. The question then is: Can the data of $g_1^p(x)$
be explained solely by $\Delta u_v(x),~\Delta d_v(x)$ and $\Delta G(x)$
without sea polarization? To examine this issue, we note that the gluon
polarization is subject to the constraint
\be
J(x,Q^2)=-{\alpha_s\over 6\pi}\int^1_x{dy\over y}\Delta\tilde{\sigma}^{\rm
hard}\left({x\over y},{Q^2\over \u^2}\right)\Delta G(y,Q^2),
\en
where
\be
J(x,Q^2) &=& g_1^p(x,Q^2)-{1\over 2}\int^1_y{dy\over y}\left[{4
\over 9}\Delta u_v(y,Q^2)+{1\over 9}\Delta d_v(y,Q^2)\right]  \non \\
&\times& \left[\delta\left(1-{x\over y}\right)+{\alpha_s(Q^2)\over 2\pi}\Delta
f_q\left({x\over y}\right)\right].
\en
One may ask: Apart from the positivity constraints, can one treat $\Delta
u_v(x),~\Delta d_v(x)$ and $\Delta G(x)$ as free parameters and fit them to
the measured $g_1^p(x)$? The point is that when one works in the
gauge-invariant factorization scheme, the shape of the polarized valence quark
distributions, which is factorization scheme independent, is constrained
by the positivity condition $|\Delta s(x)|\leq s(x)$, in particular in the
region $x\gsim 0.2$. Therefore, the l.h.s. of (31) is basically fixed by
the data of $g_1^p(x)$ and the phenomenological $\Delta u_v(x)$ and $\Delta
d_v(x)$ given by (24). The polarized gluon distribution can be extracted from
the Mellin transformation of (31) (for a detail of the procedure, see
Ref.[21]).

      The best least squares fit we found (see Fig.~6) for $\u\sim 1$ GeV
and $Q^2=Q^2_0=10\,{\rm GeV}^2$ is
\be
\Delta G(x,Q^2_0)=\,x^{-1.17}(1-x)^{5.33}(0.03-1.71\sqrt{x}+3.01x+43.5x^{1.5})
\en
with $\chi^2/{\rm d.o.f.}=10.4/22$ and $\Delta G=2.51\,$. There are two salient
features with this spin-dependent gluon density: (i) $\Delta G(x)$ is negative
at very small $x$, $x<0.025$\,. This is because the best $\chi^2$ fit to $J(x)$
is positive at small $x$. We find that $\Delta G(x)=0$ corresponds to a
maximum $xJ(x)$ occurred at $x\sim 0.025\,$. Consequently, a negative behavior
of  $\Delta G(x)$ at very small $x$ is natural. (ii) the positivity constraint
$|\Delta G(x)|\leq G(x)$ is violated at $x>0.15$ (see Fig.~6).
Hence, this $\Delta G(x)$ is physically unacceptable. However, we note
that a fit to the $g_1^p(x)$ data with the polarized gluon distribution (27),
which does respect the positivity condition, is equally acceptable with
$\chi^2/{\rm d.o.f.}=14.13/22$ (see the thick solid curve in Fig.~7). We thus
conclude that it is still
possible to reproduce the data of $g_1^p(x)$ with anomalous gluonic
contributions (of course, the shape of the gluon spin distribution is
basically arbitrary), yet a best least $\chi^2$ fit to data with $\chi^2/{\rm
d.o.f.}=10.4/22$ demands a polarized gluon distribution violating the
positivity constraint. Needless to say, we have to await high-quality data
in the future to pin down the issue.

   There exist in the literature various parametrizations for polarized
parton distribution functions fitted to the data within the framework of the
chiral-invariant factorization scheme. However, most of them are not
reliable or trustworthy owing to the incorrect use of the hard cross
section $\Delta\tilde{\sigma}^{\rm hard}$ for photon-gluon scattering, among
other things. For example, the predicted
$g_1^p(x)$ using (26) for $\Delta q_v(x)$, (27) for $\Delta G(x)$
together with the delta kernel $\Delta\tilde{\sigma}^{\rm hard}(z)=\delta
(1-z)$ fits the data very well with $\chi^2/{\rm d.o.f.}=11.9/22$ (see
the solid curve in Fig.~7). But the same set of parton spin distributions
fails to fit the data at small $x$, $x<0.01$, when the correct kernel (14) is
employed (shown by the dotted curve in Fig.~7 with $\chi^2/{\rm d.o.f.}=18.3/
22$). This is because gluon contributions at small $x$
gain more weight via the convolution with the non-delta kernel.
Recall that the same set of polarized valence
quark distributions also leads to an unacceptable sea polarization when
fitted to the data (see Sec.~III). Therefore, we believe that our valence
quark spin distributions parametrized by (24) are more sensible than
any others.

   Since in a realistic case it is likely that sea polarization is
nonvanishing and the value of $\Delta G$ is between 0 and 2.5\,, this
leads to the more realistic case (iv) $\Delta s'(x)\neq 0$ and
$\Delta G(x)\neq 0$. If we assume that the shape of $\Delta G(x)$ remains
the same as that of (33), the positivity condition of the gluon distribution
requires that
\be
\Delta G(x,Q^2_0)=\,0.199x^{-1.17}(1-x)^{5.33}(0.03-1.71\sqrt{x}+3.01x+43.5
x^{1.5}),
\en
corresponding to $\Delta G=0.5\,$. Substituting this into Eq.(30)
determines $\Delta s'(x)$, which we find can be parametrized as
\be
\Delta s'(x,Q^2_0)=-x^{-1.17}(1-x)^{9.63}(0.01\sqrt{x}+0.69x-0.949x^{1.5})
\en
with $\Delta s'=-0.087$ and $\chi^2/{\rm d.o.f.}=11.8/22\,$.

\vskip 0.3cm
\noindent{\bf V.~Discussions}
\vskip 0.3cm
  In Sections III and IV we have considered four different cases for
polarized parton distributions in the gauge-invariant and chiral-invariant
factorization schemes: (i) $\Delta s(x)\neq 0,~\Delta G(x)=0$, (ii)
$\Delta s(x)\neq 0,~\Delta G(x)\neq 0$, (iii) $\Delta s'(x)= 0,~\Delta
G(x)\neq 0$, and (iv) $\Delta s'(x)\neq 0,~\Delta G(x)\neq 0$.
Since the value of $\Delta G$ ought to lie somewhere between 0 and 2.5\,,
it appears to us that case (ii) or case (iv) is more realistic.
Note that cases (ii) and (iv) are not totally independent. This is because
for a given $\Delta G(x)$, which is factorization scheme
independent, the sea quark spin distributions $\Delta q'_s(x)$ and $\Delta
q_s(x)$ are not independent and they are related via [21] (see also Eq.(57) of
[16])
\be
\Delta q'_s(x)=\,\Delta q_s(x)+{\alpha_s\over \pi}\int^1_x{dy\over y}\left(1
-{x\over y}\right)\Delta G(y),
\en
derived from Eqs.(1), (11) and (14), where the assumption $\Delta q'_v(x)=
\Delta q_v(x)$ has been made (see Sec.~IV). Clearly, its first moment is
precisely Eq.(17), as it should be.
We have explicitly checked that $\Delta s(x)$ of (29) and
$\Delta s'(x)$ of (35) do satisfy the relation (36). The spin-dependent
parton distributions in this work are presented in the $\overline{\rm MS}$
scheme in the next-to-leading order of QCD (for a similar work, see [34]).

    It is straightforward to compute the polarized structure functions
$g_1^n(x)$ of the neutron, $g_1^d(x)$ of the deuteron and their first moments
$\Gamma_1^n$ and $\Gamma_1^d$ respectively. The various polarized
distributions satisfy the relation
\be
g_1^p(x)+g_1^n(x)=\,{2\over 1-1.5\omega_D}\,g_1^d(x),
\en
with $\omega_D=0.058$ being the probability that the deutron is in a $D$
state. We find
\be
\Gamma_1^n=-0.053\,,~~~\Gamma_1^d=0.040\,,~~~~~~{\rm at~} Q^2=10\,{\rm GeV}^2,
\en
while experimentally [1-6],
\be
\Gamma_1^n=\cases{-0.022\pm 0.007\pm 0.009\,, &E142~~at~$\langle Q^2\rangle
=2\,{\rm GeV}^2$,  \cr   -0.037\pm 0.008\pm 0.011\,, &E143~~at~$\langle Q^2
\rangle=3\,{\rm GeV}^2$,  \cr -0.063\pm 0.024\pm 0.013\,, &SMC~~at~$\langle
Q^2\rangle=10\,{\rm GeV}^2$,  \cr}
\en
and
\be
\Gamma_1^d=\cases{0.034\pm 0.009\pm 0.006\,, &SMC ~~at~$\langle Q^2\rangle
=10\,{\rm GeV}^2$,  \cr   0.042\pm 0.003\pm 0.004\,, &E143~~~at~$\langle
Q^2\rangle=3\,{\rm GeV}^2$.   \cr}
\en
Shown in Figs.~8 and 9 are the predicted polarized structure functions
$xg_1^n(x)$ and $xg_1^d(x)$ respectively at $Q^2=10\,{\rm GeV}^2$ using
the parton spin distributions in cases (i) and (ii). We see that although
our predictions are consistent in gross with experiments, new measurements
of $g_1^n(x)$ and $g_1^d(x)$ with refined accuracy are certainly needed.
Note that the $Q^2$ dependence of polarized structure functions is not
discussed here since our parton spin distributions are parametrized at $Q^2_0=
10\,{\rm GeV}^2$ and it is known that only $Q^2>Q^2_0$ evolution is
governed by the Altarelli-Parisi equations.

\vskip 0.3 cm
\noindent{\bf VI.~Conclusions}
\vskip 0.3cm
The fact that the size of the hard-gluonic contribution to $\Gamma_1^p\equiv
\int^1_0g_1^pdx$ is purely a matter of the factorization convention chosen
in defining the quark spin distribution promotes us to consider four
different possibilities of polarized parton distribution functions in two
extreme factorization schemes: gauge-invariant and chiral-invariant ones.
One cannot tell experimentally whether or not gluons contribute to
$\Gamma_1^p$. We stressed that
the hard cross section for photon-gluon scattering is unique up to the
factorization scale $\u$ and is independent of the choice of the soft and
ultraviolet regulators.

     Owing to the positivity constraints for sea and gluon polarizations,
$g_1^p(x)$ at $x\gsim 0.2$ should receive almost all contributions from
polarized valence quark distributions. This together with the first moment
and perturbative QCD constraints puts a very nice restriction on the shape
of $\Delta u_v(x)$ and $\Delta d_v(x)$. Eq.(24) is our best result for valence
quark spin distributions at average $\la Q^2\ra=10\,{\rm GeV}^2$.
Working in the gauge-invariant factorization scheme,
we have extracted the polarized sea distribution function from the EMC and
SMC data of $g_1^p(x)$ with the results (25) and (29) for cases (i) and (ii)
respectively. All polarized parton distributions in this work are presented
in the next-to-leading order of QCD at the scale $Q^2=10\,{\rm GeV}^2$.

  Based on the chiral invariant scheme and the aforementioned valence
quark spin densities, we have found that it is possible to explain the
measurements of $g_1^p(x)$ with anomalous gluonic contributions,
yet a least $\chi^2$ fit to
the data indicates that the best fitted gluon spin distribution violates
the positivity condition $|\Delta G(x)|\leq G(x)$. We have considered
a more realistic set of parton spin distributions with a moderate value
of $\Delta G=0.5\,$ and with a nonvanishing sea polarization. Many
parametrizations of polarized parton distributions presented in the literature
are not trustworthy due mainly to the use of an incorrect hard cross section
for photon-gluon scattering.

  In principle, the choice of the set of $\Delta q(x),~\Delta G(x),~
\Delta\sigma^{\rm hard}(z)$ or of $\Delta q'(x),~\Delta G(x),~\Delta\tilde{
\sigma}^{\rm hard}(z)$ to describe the polarized hadron structure function
is a matter of convention. In fact, for a given $\Delta G(x)$, $\Delta
q'(x)$ and $\Delta q(x)$ are related via Eq.(36). In practice, the
gauge-invariant quantity $\Delta q$ is probably more convenient and natural
to use since it can be
expressed as a nucleon matrix element of a local gauge-invariant operator.
It is calculable in lattice QCD and, more importantly, its $Q^2$ evolution
is directly governed by the polarized Altarelli-Parisi equations, which is
not the case for $\Delta q'$ and $\Delta q'(x)$ [see the footnote after
Eq.(15)].

    Of course, inclusive polarized deep inelastic scattering experiments alone
cannot reveal the magnitude and shape of the gluon spin distribution, and one
has to await measurements of $\Delta G$ in independent processes in order to
fully understand the proton spin structure. Nevertheless, there does exist
a truly theoretical progress since the EMC measurement of $g_1^p$, namely
the lattice calculation of the proton matrix elements of the axial current
[8,9]. The empirical SU(3) invariance observed by lattice QCD for the sea
polarization manifested in the disconnected insertion strongly suggests that
it is the
axial anomaly which is responsible for the negative sea polarization and which
explains the smallness of the quark spin content of the proton.

\vskip 2.5 cm
\centerline{\bf ACKNOWLEDGMENTS}
\vskip 0.7 cm
   One of us (H.Y.C.) wishes to thank K.F. Liu and X. Ji for many helpful
discussions.
    This work was supported in part by the National Science Council of ROC
under Contract No. NSC85-2112-M-001-010.

\pagebreak
\vskip 0.8 cm
\centerline{\bf REFERENCES}
\vskip 0.3 cm
\begin{enumerate}

\item SMC Collaboration, D. Adams {\it et al.,} \pl {\bf B329}, 399 (1994);
{\bf B339}, 332(E) (1994).

\item E143 Collaboration, K. Abe {\it et al.,} \prl {\bf 74}, 346 (1995).

\item E142 Collaboration, D.L. Anthony {\it et al.,} \prl {\bf 71}, 959
(1993).

\item SMC Collaboration, B. Adeva {\it et al.,} \pl {\bf B302}, 533 (1993);
D. Adams {\it et al.,} \pl {\bf B357}, 248 (1995).

\item E143 Collaboration, K. Abe {\it et al.,} \prl {\bf 75}, 25 (1995).

\item EMC Collaboration, J. Ashman {\it et al.,} \np {\bf B238}, 1 (1990); \pl
{\bf B206}, 364 (1988).

\item R. Gupta and J.E. Mandula, \pr {\bf D50}, 6931 (1994); R. Altmeyer,
M. G\"ockler, R. Horsley, E. Laermann, and G. Schierholz, \pr {\bf D49}, 3087
(1994); B. All\'es, M. Campostrini, L. Del Debbio, A. Di Giacomo, H.
Panagoulos, and E. Vicari, \pl {\bf B336}, 248 (1994).

\item S.J. Dong, J.-F. Laga\"e, and K.F. Liu, \prl {\bf 75}, 2096 (1995).

\item M. Fukugita, Y. Kuramashi, M. Okawa, and A. Ukawa, \prl {\bf 75}, 2092
(1995).

\item J. Ellis and R. Jaffe, \pr {\bf D9}, 1444 (1974).

\item G.T. Bodwin and J. Qiu, \pr {\bf D41}, 2755 (1990), and
in {\it Proc. Polarized Collider Workshop}, University Park, PA, 1990, eds.
J. Collins {\it et al.} (AIP, New York, 1991), p.285.

\item P. Ratcliffe, \np {\bf B223}, 45 (1983).

\item R.D. Carlitz, J.C. Collins, and A.H. Mueller, \pl {\bf B214}, 229
(1988).

\item G. Altarelli and G.G. Ross, \pl {\bf B212}, 391 (1988).

\item R.L. Jaffe and A.V. Manohar, \np {\bf B337} 509 (1990).

\item S.D. Bass and A.W. Thomas, {\sl J. Phys.} {\bf G19}, 925 (1993);
Cavendish preprint 93/4 (1993).

\item A.V. Manohar, \prl {\bf 66}, 289 (1991).

\item A.V. Manohar, in {\it Proc. Polarized Collider Workshop}, University
Park, PA, 1990, eds. J. Collins {\it et al.} (AIP, New York, 1991), p.90;
see also R.D. Carlitz and A.V. Manohar, {\sl ibid.} p.377.

\item Particle Data Group, \pr {\bf D50}, 1173 (1994).

\item J. Ellis and M. Karliner, \pl {\bf B341}, 397 (1995).

\item H.Y. Cheng and C.F. Wai, \pr {\bf D46}, 125 (1992).

\item G.R. Farrar and D.R. Jackson, \prl {\bf 35}, 1416 (1975);
S.J. Brodsky, M. Burkardt, and I. Schmidt, \np {\bf B441}, 197 (1995).

\item A.D. Martin, R.G. Roberts, and W.J. Stirling, \pr {\bf D50}, 6734
(1994); \pl {\bf B354}, 155 (1995).

\item NMC Collaboration, P. Amaudruz {\it et al.,} \pl {\bf B295}, 159 (1992).

\item D.J.E. Callaway and S.D. Ellis, \pr {\bf D29}, 567 (1984).

\item T. Gehrmann and W.J. Stirling, \zp {\bf C65}, 461 (1995).

\item G. Preparata and J. Soffer, \prl {\bf 61}, 1167 (1988);
{\bf 62}, 1213(E) (1989).

\item J. Soffer, CPT-92-P-2809 (1992).

\item S.A. Rabinowitz {\it et al.,} \prl {\bf 70}, 134 (1993).

\item G. Altarelli and W.J. Stirling, {\sl Particle World} {\bf 1}, 40
(1989).

\item J. Ellis, M. Karliner, and C.T. Sachrajda, \pl {\bf B231}, 497 (1989).

\item G.G. Ross and R.G. Roberts, RAL-90-062 (1990).

\item A.V. Efremov and O.V. Teryaev, in {\it Proceedings of the International
Hadron Symposium}, Bechyne, Czechoslovakia, 1988, eds. Fischer {\it et al.}
(Czechoslovakian Academy of Science, Prague, 1989), p.302.

\item M. Gl\"uck, E. Reya, M. Stratmann, and W. Vogelsang, DO-TH 95/13,
RAL-TR-95-042 (1995); R.D. Ball, S. Forte, and G. Ridolfi, \np {\bf B444},
287 (1995); CERN-TH/95-266 (1995).

\end{enumerate}
\pagebreak

\centerline{\bf FIGURE CAPTIONS}
\vskip 1.0cm
\begin{description}

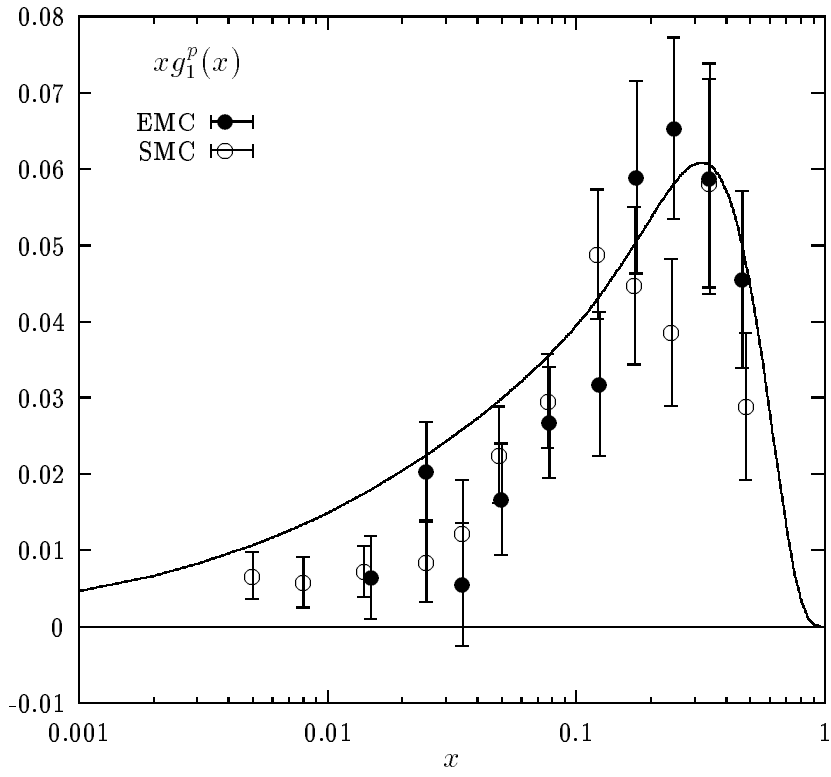
\item[Fig.~1] The theoretical curve of $xg_1^p(x)$ fitted to the EMC and SMC
data at $x\gsim 0.2$ with the polarized valence quark distributions given
by (24) and without sea and gluon polarizations.

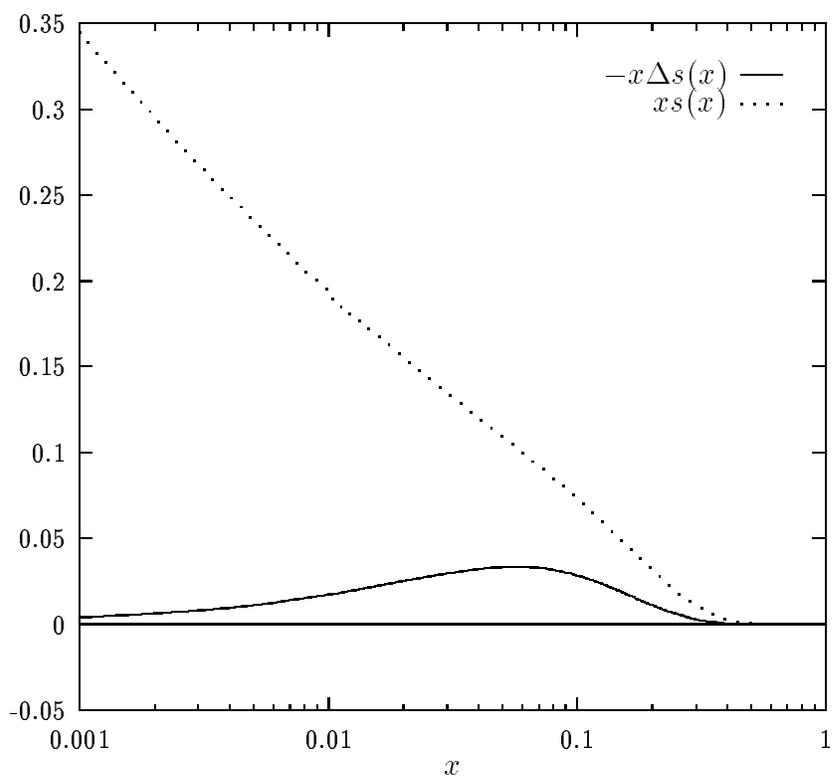
\item[Fig.~2] The polarized strange quark distribution $-\Delta s(x)$
fitted to the data of $g_1^p(x)$. Also shown is the unpolarized strange quark
distribution evaluated at $Q^2=10\,{\rm GeV}^2$ using the MRS(A$'$)
parametrization [23].

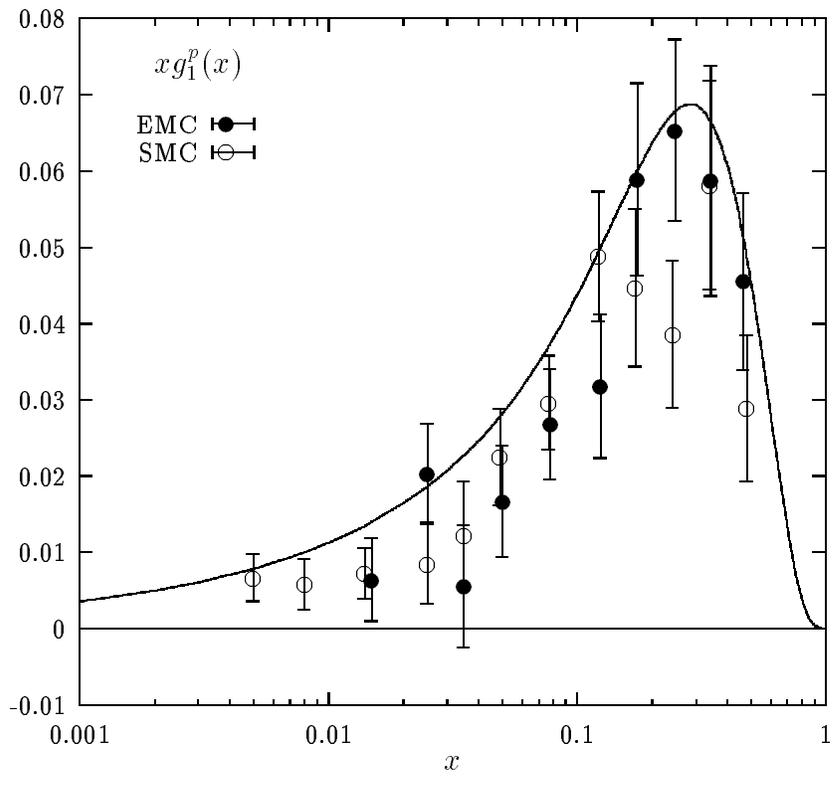
\item[Fig.~3] The predicted curve of $xg_1^p(x)$ arising from the
spin-dependent valence quark spin distributions given by (26) without sea
and gluon contributions. At first sight, it appears to give a reasonable
eye-fit to the data.

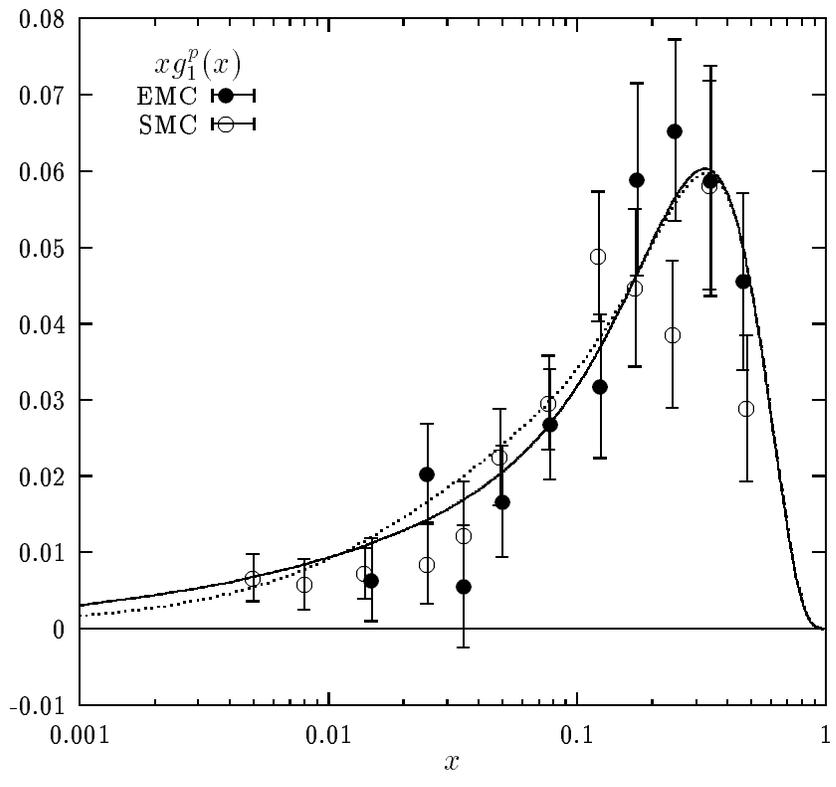
\item[Fig.~4] Two theoretical curves for $xg_1^p(x)$. The solid line is
the predicted curve for case (i) with $\chi^2/{\rm d.o.f.}=12.24/22$, and
the dotted curve with $\chi^2/{\rm d.o.f.}=14.95/22$
is for case (i) plus the polarized gluon distribution
given by (27).

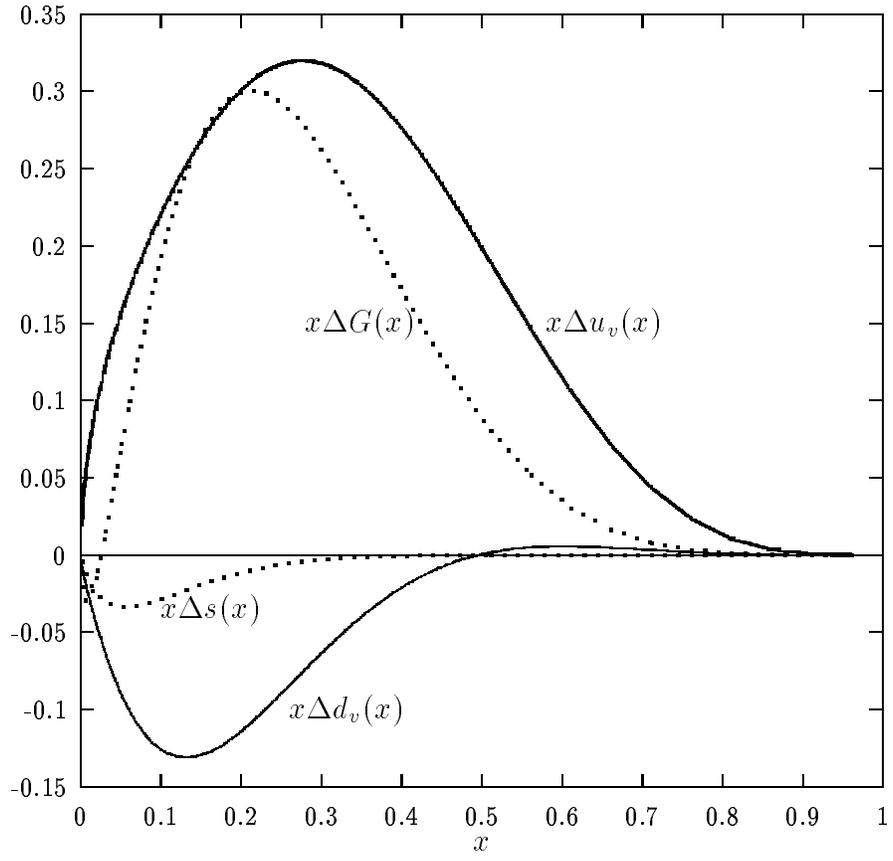
\item[Fig.~5] Parton spin distributions for case (ii) parametrized at
$Q^2=10\,{\rm GeV}^2$.

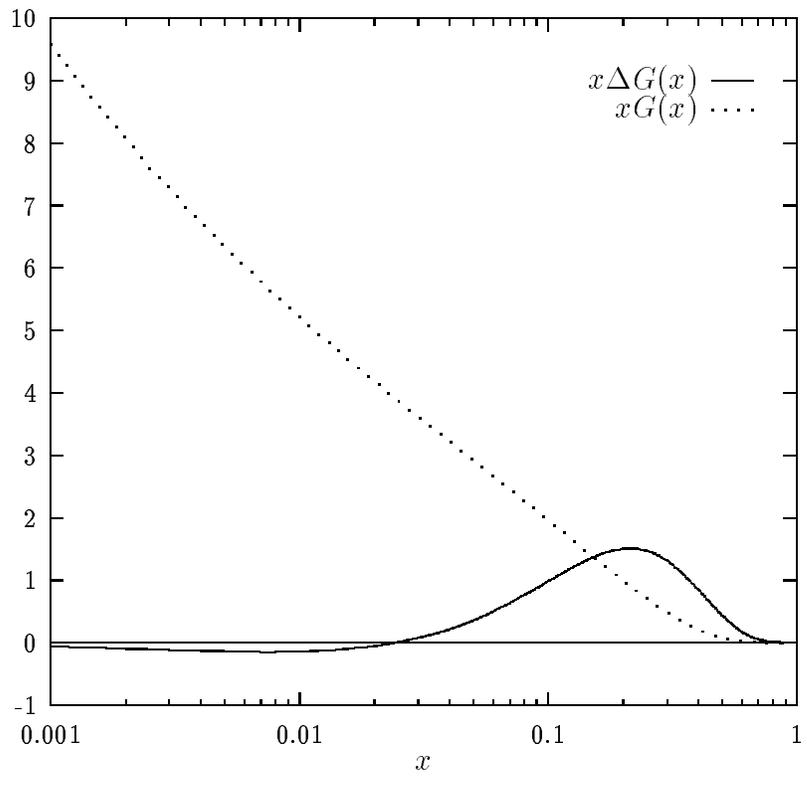
\item[Fig.~6] The polarized gluon distribution extracted from a best least
$\chi^2$ fit to the data of $g_1^p(x)$ by assuming $\Delta s'(x)=0$.
Also shown is the unpolarized gluon distribution evaluated at
$Q^2=10\,{\rm GeV}^2$ using the MRS(A$'$) parametrization [23].

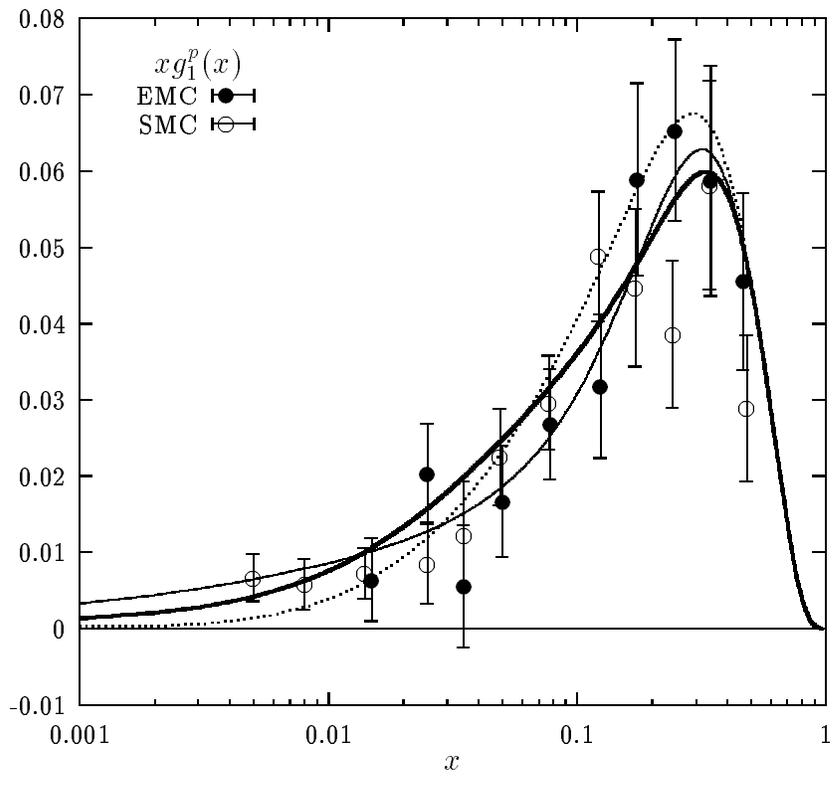
\item[Fig.~7] Three theoretical curves for $xg_1^p(x)$.
With (27) for the polarized gluon distribution, the thick solid curve
is calculated using (24) for valence quark spin distributions and (14) for
the kernel, while the solid and dotted curves are based on (26) for
$\Delta q_v(x)$ and the delta kernel $\Delta\sigma(z)=\delta(1-z)$ for the
former curve and the kernel (14) for the latter.

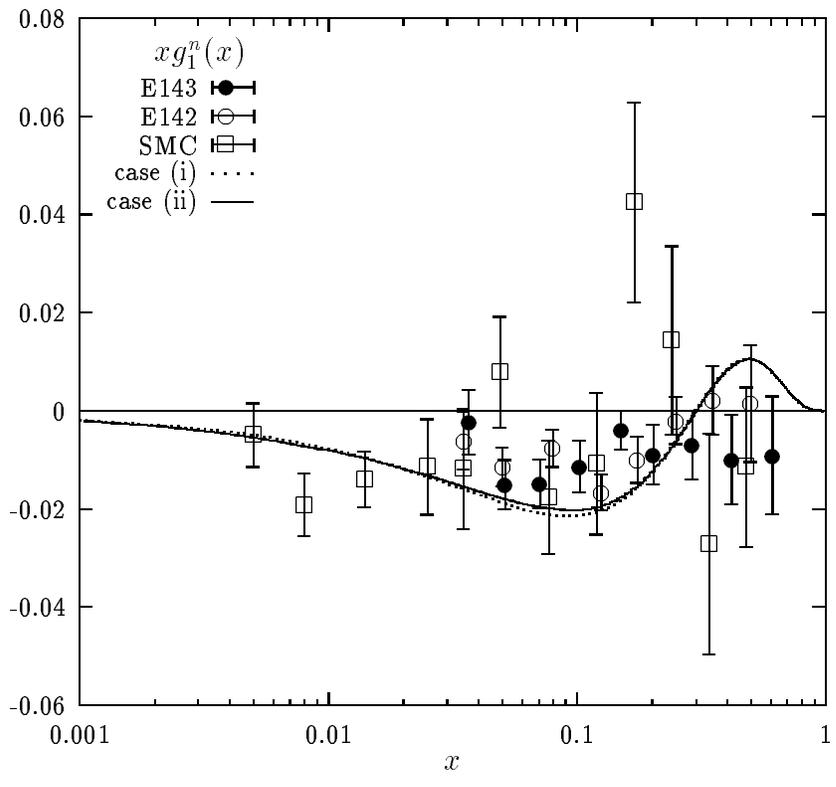
\item[Fig.~8] The predicted polarized structure function $g_1^n(x)$ of the
neutron at $Q^2=10\,{\rm GeV}^2$ using the parton spin distributions
in cases (i) and (ii). Also shown are the E142, E143 and SMC data at
the average $Q^2$ of each $x$ bin.

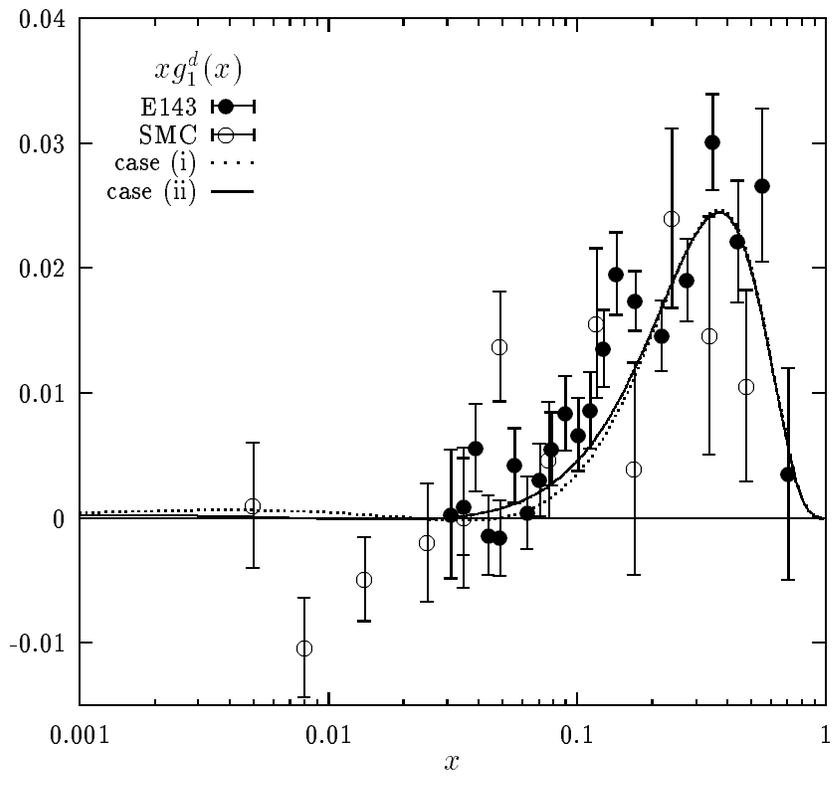
\item[Fig.~9] Same as Fig.~8 except for the deuteron. The SMC data of
$g_1^d(x)$ are evaluated at $Q^2=10\,{\rm GeV}^2$, while E143 data at the
average $Q^2$ of each $x$ bin.

\end{description}
\end{document}